# Optimized Routing and Spectrum Assignment for Video Communication over an Elastic Optical Network

H. Alizadeh Ghazijahani, H. Seyedarabi, J. Musevi Niya, and N. M. Cheung

*Abstract*—Elastic optical network (EON) efficiently utilize spectral resources for optical fiber communication by allocating minimum necessary bandwidth to client demands. On the other hand, network traffic has been continuously increasing due to the wide penetration of video streaming services, so the efficient and cost-effective use of available bandwidth plays an important role in improving service provisioning. In this work we formulate and solve an optimization problem to perform routing and spectrum assignment (RSA) in EON with focus on video streaming. In this formulation, EON and video constraints such as spectrum fragmentation and received video quality are considered jointly. In this way, we utilize a machine learning (ML) technique to estimate the video quality versus channel state. The proposed algorithm is evaluated over two benchmark fiber optic network, namely NSFNET and US-backbone using numerical simulations based on random traffic models. The results reveal that the mean optical signal-to-noise ratio (OSNR) for video content data in the receiver is remarkably higher than in non-video data. This is while the blocking ratio is the same for both data types.

## I. INTRODUCTION

ALONG with the increasing bandwidth-intensive multimedia applications such as 3D video-on-demand, e-learning, high-definition television, etc., video data communication is being grown [1,2]. It is expected by 2021, 80% of the world's internet traffic will be video [3]. The increase in data traffic demand leads to congestion in fiber optic data links. Traditional wavelength division multiplexing (WDM) based network operates within fixed frequency grids by assigning a fixed wavelength channel to a connection request. In this scheme, all demands receive the same bandwidth based on the rigid spectral grids defined by ITU standard. Consequently, often the most part of the assigned bandwidth remains unused. Recently, elastic optical network has been proposed to handle heterogeneous traffic by allocating flexible spectrum resources referred to as frequency slots (FSs) [4]. Utilizing coherent optical orthogonal frequency division multiplexing (CO-OFDM), EON provides high data rate beyond 100 GB/s, while achieving higher spectrum resource efficiency [5,6].

One of the basic concepts in EON is routing and spectrum assignment (RSA) problem. The goal of RSA is to find the best route and establish a light-path in an all-optical channel for the given end to end demands. RSA is an NP-hard optimization problem where some constraints should be considered to reach an optimal efficiency of available spectrum. Solving RSA is a challenging subject and some works have been presented in the literature. In [7] a fragmentation-aware RSA scheme in EON is introduced which evaluate the accommodation capability of spectrum resources for the incoming request before spectrum assignment. Moharrami *et al.* [1] formulate an integer linear programming (ILP) to perform multicast routing and spectrum assignment, where select paths with regard to network congestion. Furthermore, some studies addressing distance adaptive RSA [8-10] in which that make a tradeoff among bitrate, bit error rate (BER) and modulation format spectral efficiency. Yaghubi *et al.* [11] proposed a method to find a near optimal solution for the joint RSA and modulation level determination problem. They have exploited stepwise greedy algorithm to solve the formulated ILP problem. Authors in [12] focus on power influences on physical layer effects and solve a PRMLSA (Power, Routing, Modulation Level and Spectrum Assignment) problem.

High Efficiency Video Coding, known as HEVC, is the most recent joint video project of the ITU-T video coding experts group (VCEG) and the ISO/IEC moving picture experts group (MPEG) standardization organizations, working together in a partnership known as the joint collaborative team on video coding (JCT-VC) [13]. One of the most attractive areas of multimedia streaming is wireless multimedia sensor networks

Hamed Alizadeh Ghazijahani is with the faculty of electrical and computer engineering, University of Tabriz, Iran. (e-mail: hag@tabrizu.ac.ir).

Hadi Seyedarabi (corresponding author) is with the faculty of electrical and computer engineering, University of Tabriz, Iran. (e-mail: seyedarabi@tabrizu.ac.ir).

Javad Musevi Niya is with the faculty of electrical and computer engineering, University of Tabriz, Iran. (e-mail: niya@tabrizu.ac.ir).

Ngai-Man Cheung is with the information systems technology and design pillar, Singapore university of technology and design, Singapore. (e-mail: ngaiman_cheung@sutd.edu.sg).

(WMSN) where enable new applications such as multimedia surveillance, advanced control systems, health care delivery and monitoring. According to the application constraints, WMSN tries to provide the best quality-of-experience (QoE) regarding delay and quality of the received video. Most of the works have been focused on access and aggregation layers of wireless networking while having the concern on limited battery life, bandwidth and other channel limitations [14,13,15-17]. This is while, to the best of our knowledge, there is no any study addressed video streaming over EON as a backbone network and beyond challenges.

The future of internet data handling backbone network is EON, accordingly, due to the time-varying network conditions, especially traffic load, an efficient RSA scheme is needed to support streaming of large video data considering its constraints. In this work, we present an efficient routing and spectrum assignment scheme for EON guaranteeing user's video QoE. In this scheme, the congestion and the physical conditions of network links are considered besides video quality metrics.

Some models have been presented in literature estimate video quality versus channel state. A model introduced in [18] calculates frame error rate versus packet error rate, but not model presented yet to show the functionality of frame error rate with channel status, e.g. BER. Furthermore, Dalei *et al.* in [19] use the presented sigmoid function-based model in [20] for packet error probability estimation where is not specialized for video data packet format. We need a model that gives an estimation of decoded video quality after passing the video signal through a communication channel. Subsequently, the available models did not satisfy our area.

In this way, we utilize *newRB* as a machine learning (ML) method to estimate the decoded video QoE metrics, namely peak signal-to-noise ratio (PSNR) and decodable frames rate (DFR) versus channel state. In our problem, the network is EON and the communication channel is fiber optic. To calculate the optical signal-to-noise ratio (OSNR) of a path we have utilized the provided model by Gao, *et al.* [21], that derived an analytical expression for the noise spectral density included by the fiber nonlinearities for a multichannel CO-OFDM system.

To solve the proposed RSA problem, we considered an optimization problem where the cost function includes both network and video costs. The network cost is determined by the efficiency of the spectrum resources utilization and the video cost is a function of decoded video QoE metrics.

The rest of this paper is organized as follows. Section 2 analyses the fragmentation and misalignment problems in EON and overviews video traffic types from the application point of view. Section 3 is devoted to describing the system model. The proposed RSA algorithm is introduced in section 4. The algorithm evaluation and the simulation results are presented in section 5. Finally, section 6 concludes the paper.

## II. PROBLEM STATEMENT

The advantages of EON make it as the promising technology for optical core networks. These advantages of EON are to facilitate the applications to provide high-quality experiences to the users. For the internet video applications, the received quality is influenced by control decisions taken at each and every layer of the protocol stack. Here, the challenge is to solve the RSA problem at the network layer for a video delivery request subject to experience a satisfactory video quality besides high efficient use of spectrum in EON.

### A. Video Traffic

Video traffic has boundaries according to the application that must be considered to satisfy the quality of received video. The most common parameters to evaluate the video QoE at the receiver side contains delay, PSNR of received frames, and DFR. In real-time video which the video is being used for some real-time applications such as video conference, delay is of great importance. So that, even small delays can have a significant impact on the quality of the video communication experience. Video gaming is almost similar to online applications, where latency is the primary concern. But in source coding point of view, due to the high motion nature of sequences, the compression rate may be lower than in online slow-motion video. As a result, considering the same PSNR, video gaming traffic needs more bandwidth than online video conference, consequently this type of video forces more traffic load to the network. Unlike real-time video services, video on demand services transmit pre-recorded content based on the demands of the end user. The end user waiting for a higher quality of such services than online ones. So, the network management system should give two quality factors in priority, namely PSNR and DFR. Interactive video is a type of video on demand service, which provides playback capability to the user. This includes traditional playback commands such as pause, fast forward, and rewind. Internet video applications such as YouTube and Netflix are the origin of interactive video service. Another type of video services is multimedia surveillance used to monitor an area. Such services provide real-time video to an end user. The delivered video must have an acceptable resolution with high PSNR so that an unauthorized or unexpected activity can be detectable. Having low latency and high PSNR is of most importance in such applications [24].

### B. Elastic Optical Network

The traditional WDM optical network follows the rigid ITU-T spectrum partitioning standard and divides the spectrum to fixed grid channels by 50 GHz or 100 GHz. In WDM a fixed wavelength is assigned to a request where, if a demand carries low bandwidth, the residual bandwidth will be lost and wasted [22]. To solve the spectrum mismatching, the EON divides the spectrum into finer bandwidths and assigns nearly exact number of frequency slots to a connection request. The OFDM based EON has the ability to support heterogeneous traffic demands in a much more spectrum-efficient way than WDM networks [23].

At the networking level, the EON control plane needs to find the routing path and allocate a sufficient number of successive FSs to a connection request along the path from source to destination. This function is called routing and spectrum assignment, in brief RSA, so that it must consider two constraints, namely spectrum continuity, and spectrum contiguity. According to the spectrum continuity, each light path utilizes the same spectrum slots in the intermediate connection links through its path. The spectrum contiguity dictates the system to assign successive spectrum slots to each demand. However, these two constraints will lead to fragmentation in the spectrum, which would increase traffic blocking in the network.

## III. System Model and Definitions

In this system we try to calculate the end-to-end OSNR based on the considered route in the network and get the BER. Employing neural network presents an estimation of PSNR and DFR versus given BER. Accordingly, we formulate an optimization problem that minimizes the spectrum fragmentation in the EON and maximizes the video QoE.

### A. EON Model and Constraints

The EON is attractive due to its nature of flexibility in data rate and spectrum allocation. Two key components which provide elasticity for EON are bandwidth-variable transponder (BVT) and bandwidth-variable wavelength cross-connect (BV-WXC). The functionality of BVTs is to tune the client data signal by allocating just enough bandwidth. Concurrently, the intermediate BV-WXCs are used to add and drop local signals and switching and routing for transit signals.

There are some metrics defined in literature to measure the spectrum fragmentation (e.g. link fragmentation ratio, possible accommodation states) [7]. In this work, we use the fragmentation cost parameters defined by Yin *et al.* [25], namely number of *cuts* and *misalignments*. The *cuts* are considered to be the costs of the candidate solutions, since more cuts create more fragments on the candidate links of the routes. The *misalignment* counts the increase or decrease of the fragmentation between the neighbouring links. Less *cuts* and less *misalignments* brings opportunity for the optical network to have more respectively contiguous and continuous FSs to accommodate the coming connection requests in the future [25,26]. To express these concepts, consider a connection request needs two FSs to establish from source A to destination F in the given network shown in Fig. 1-a. Considering shortest path routing scenario, there are two candidate routes from node A to node F. Rout #1 and Rout #2 use Link2, Link5, Link8 and Link1, Link4, Link7, respectively to handle the traffic. Considering 10 FSs for each link, the status table of the network is shown in Fig. 1-b, where the white blocks are the free spectrum-spatial blocks and in use ones are shown with grey blocks. Considering continuity constraint in route ACEF, two contiguous FSs from the collection of {8, 9, 10} can be chosen. If one chooses FSs = {8, 9}, the Link 2 and Link 8 will be fragmented so the *cut* = 2. By this selection the misalignments are: <Link1, Link2> = +1−1 = 0, <Link3, Link2> = 2, <Link3, Link5> = 2, <Link6, Link5> = 2, <Link6, Link8> = 2, <Link7, Link8> = −2, so the total *misalignment* will be 6. On the other hand, if one chooses FSs = {9, 10} for candidate route ACEF, the *cuts* will be 0 and misalignments as: <Link1, Link2> = 2, <Link3, Link2> = 2, <Link3, Link5> = 2, <Link6, Link5> = 2, <Link6, Link8> = 2, <Link7, Link8> = −2, so the total *misalignment* will be 8. About candidate route ABDF, only FSs = {5, 6} are common free contagious ones available in Link1, Link 4 and Link 7. The only cut happens in Link 4. By this selection the misalignments are: <Link2, Link1> = 2, <Link3, Link1> = −2, <Link3, Link4> = −2, <Link6, Link4> = -2, <Link6, Link7> = -2 so the total *misalignment* for this route will be −6. Considering the number of cuts and neighbour misalignments as the costs for each candidate route and FS sets, it is wise to select Route #2 and assign FSs = {5, 6} to establish the demanded traffic.

### B. Quality of Transmission Estimation in Optical Fiber

In fiber optic networks, factors that affect OSNR are amplified spontaneous emission (ASE) noise produced by spontaneous emission in amplifiers and nonlinearities such as four-wave mixing, cross-phase modulation, and self-phase modulation. Indeed, each BV-WXCs in the routing nodes impose degradation on OSNR [27,28]. We need a closed-form expression for OSNR to estimate the quality of transmission of each elastic spectrum path in EON. In [21] Gao, *et al.*, derived an analytical expression for noise spectral density ($I_{NL}$) included by the fiber nonlinearities for a multichannel CO-OFDM system. They considered 2$N$+1 CO-OFDM channels (here FSs)

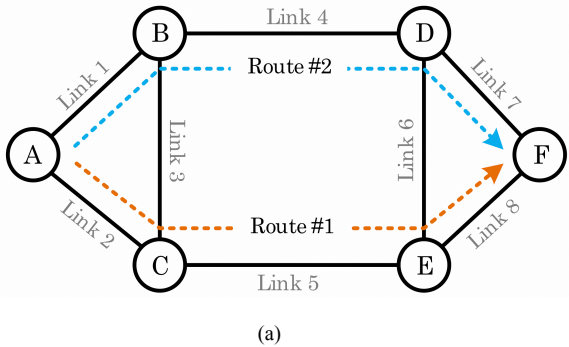 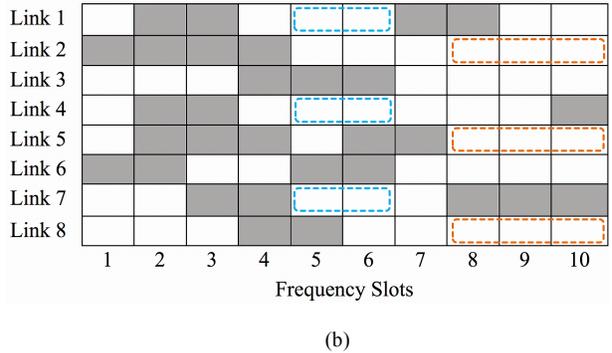

(a)          (b)

Fig. 1. a) A sample network with two candidate routing paths from source node A to destination node F; b) The status table shows the free spectrum-spatial blocks by white and in use blocks by grey.

with equal bandwidth of $B_1$ and obtained $I_{NL}$ as follows

$$I_{NL} = \left(\frac{I}{I_0}\right)^2 I$$

$$I_0 \approx \sqrt{\frac{\pi\alpha|\beta_2|}{\gamma^2 N_s h_e}} \left(1 - \frac{\Delta g}{\Delta B}\right)^{-0.5} \left[\ln\left(\frac{B}{B_0}\right) - \frac{\Delta g}{\Delta B}\ln(2N-1)\right]^{-0.5} \quad (1)$$

where the optical fiber physical parameters $\gamma$, $\alpha$, and $\beta_2$ are the nonlinearity coefficient, attenuation factor, and group velocity dispersion parameter, respectively. In addition, $N_s$ is the number of fiber spans, $I$ is the spectral density of launched power and $h_e$ is the nonlinear multi-span noise enhancement factor given by [27]

$$h_e = \frac{2\left(N_s - 1 + e^{-\alpha\varsigma L N_s} - N_s e^{-\alpha\varsigma L}\right)e^{-\alpha\varsigma L}}{N_s\left(e^{-\alpha\varsigma L} - 1\right)^2} + 1 \quad (2)$$

where $L$ is the length of each fiber span and $\zeta$ denotes the residual dispersion ratio. Furthermore, $B$ is the total fiber bandwidth, $\Delta B = B_1 + \Delta g$, $B_0 = \frac{4f_\omega^2}{B}$ and $f_\omega = \frac{1}{2\pi}\sqrt{\frac{\alpha}{|\beta_2|}}$. Considering nonlinearity in the fiber, the distribution of signal spectral density is as $I e^{-\left(\frac{I}{I_0}\right)^2}$. Hence, the OSNR in the presence of nonlinearity and ASE noises is obtained as [27]

$$\text{OSNR} = \frac{I e^{-\left(\frac{I}{I_0}\right)^2}}{n_0 + I - I e^{-\left(\frac{I}{I_0}\right)^2}} \cong \frac{I}{n_0 + I\left(\frac{I}{I_0}\right)^2} \quad (3)$$

The ASE noise parameter computed as $n_0 = 0.5 N_s e^{\alpha L} h\upsilon N_F$ where, $h$, $\upsilon$, and $N_F$ are Planck constant, light frequency and noise figure of the optical amplifier, respectively [28]. The noise model of each intermediate BV-WXC can be considered as a penalty factor ($n_{XC}$). Hence, the source-destination OSNR of a path with $N_H$ hops is given in decibel by

$$\text{OSNR}_{S-D}(\text{dB}) = \text{OSNR}(\text{dB}) - N_H \times n_{XC}(\text{dB}) \quad (4)$$

The BER is calculated according to the utilized modulation format and given OSNR$_{S-D}$. On the other hand, the BER will be affected by the error correcting capability of channel coding method being used at the transmitter. The effect of channel coding can be taken into account by assuming that the virtual apparent signal to noise ratio at the receiver is as

$$\text{OSNR}_{S-D,app} = d_{free}.R_C.\text{OSNR}_{S-D} \quad (5)$$

where $d_{free}$ and $R_C$ ($< 1$) are the free distance and rate of the code, respectively [18].

### C. MPEG Video and quality metrics

In predictive video encoding, a group of picture (GOP) is defined as a collection of successive pictures within a coded video stream. There are three different frame types; Intra-coded, predictive and bidirectional frames, named P, I and B frames, respectively [29]. In this study we consider a typical GOP structure with 12 frames to be encoded as IBBPBBPBBPBB [18,30]. The video quality at the receiver suffers from distortion caused by the transmission channel, which is a function of BER. This yields some video frames to be not decodable at the receiver and some to reduction desirable quality. We define a utility function to measure the user's satisfaction level as the product of DFR and PSNR (of the decodable frames). The PSNR evaluates the objective quality of received video stream as

$$\text{PSNR} = 10\log\left(\frac{255^2}{MSE}\right)$$

$$MSE = \frac{1}{mn}\sum_{i=1}^{m}\sum_{j=1}^{n}\left[I_{i,j} - K_{i,j}\right]^2 \quad (6)$$

where $I_{i,j}$, and $K_{i,j}$ denote the value of $i, j$ pixel respectively in original and reconstructed video frames of size $m$, $n$ pixels. Moreover, DFR = $N_D/N_T$, with $N_D$ and $N_T$ represent the number of decoded frames at receiver and the total number of transmitted frames, respectively.

To implement the discussed utility function for the received video we need to estimate the BER effect on DFR and PSNR of decoded frames. There are some works estimate video quality versus channel state. Dalei *et al.* in [19] use the sigmoid function based model presented in [20] for packet error probability. The parameters used in this model are unclear and the model is not specialized for video data packet format. Furthermore, a model introduced in [18] calculates frame error rate versus packet error rate. To the best of our knowledge, not any model presented yet to show the functionality of PSNR and DFR versus BER channel.

In this work, we employed a ML technique, namely *newRB*, to estimate the video quality metrics versus BER at the application layer, shown in Fig. 2. The ground truth data is made with heavy simulations, so as a set of HEVC video files are subjected to the diverse BERs then decoded and DFR and PSNR of them are extracted. Almost 2/3 of the ground truth data is used to train and 1/3 of data is used to test the proposed neural network. Fig. 3 shows the training performance of the neural network versus training epochs.

## IV. PROTOCOL DESIGN

### A. Algorithm

The proposed RSA algorithm takes into account both the EON and video constraints. Here, the idea is to find an appropriate connection path, includes routing path and sufficient FSs, for both video and non-video data types subject to increase network spectrum efficiency and keep user QoE.

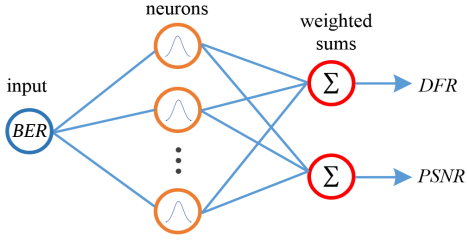

Fig. 2. The *newRB* neural network model with *BER* as input and *DFR* and *PSNR* as output parameters.

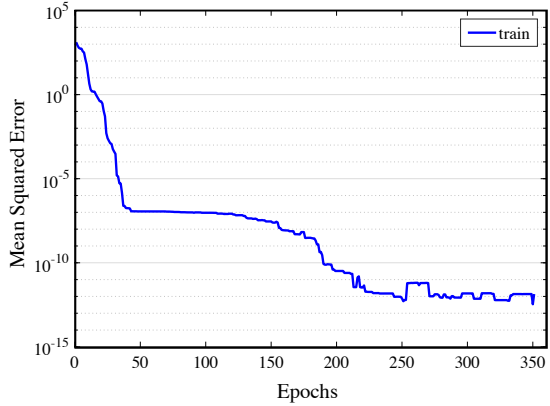

Fig. 3. Training performance of *newRB* versus epochs.

Algorithm I summarizes the proposed protocol. As request $q$ arrives, the proposed RSA algorithm finds a number of candidate paths based on a common shortest path routing algorithm, e.g. *Dijkstra*. As well, the unoccupied FSs in each link are listed. In the next step, the cost of each candidate path is computed as follows

$$F^q_{cost} = \alpha F^q_{Network} + \beta F^q_{Video} \quad (7)$$

$$F^q_{Network} = N_c + \frac{N_m}{S_q N_{NL}} \quad (8)$$

$$U(BER_q) = PSNR(BER_q) \times DFR(BER_q) \quad (9)$$

$$F^q_{Video} = \log\left(\frac{1}{U(BER_q)}\right) \quad (10)$$

with $N_c$, $N_m$ and $N_{NL}$ representing the number of the cuts, misalignments, and neighbour links in the candidate path, respectively. The number of FSs requested by session $q$ is noted by $S_q$. Larger utility value gives higher user QoE and correspondingly provides low cost, so we consider an inverse relationship for utility in video cost function. Additionally, the relatively large range of utility values encouraged us to propose logarithmic function, presented in (10). The cost function of the network, i.e. $F_{Network}$, computes the cost parameters related to the EON network where measures the fragmentation of the spectrum assuming a candidate link establishment. As well, the cost function of video, i.e. $F_{Video}$, estimates inverse of the video QoE versus given channel state. The chosen values for weighting coefficients in (7), i.e. $\alpha$ and $\beta$, determines the significance of network or video costs on total cost and influences the final decision on lightpath selection. If the content of requested connection is not video data, only the cost of the network will be considered, i.e. $\alpha = 1$, $\beta = 0$.

The routing path and FSs with the minimum cost are selected. If the selected path qualifies the video quality criterion, which is determined according to the user QoE for a session, then the source-destination connection is established. Hence, if not any link guarantees expected video quality, the request is blocked due to the qualification constraint. On the other hand, if no path is found during path computation phase, lightpath request is blocked due to congestion in the network.

### B. Optimization formulation

The proposed RSA algorithm guaranteeing video QoE proposed in Algorithm I can be presented as an optimization problem as below

**Notations:**

$N_{FS} = 2N+1$, number of FSs in each fiber link;

$E$, the connection links of the network

$S^{FS}_{l,q}$, a matrix of size $N_{FS} \times |E|$ where keeps the occupancy status of request $q$ in link $l$ *in frequency slots FS*;

$Q$, the live connections in the network

$U_{th}$, the minimum utility function value requested for a video streaming to be guaranteed;

$f_q$, the index of the central frequency of assigned bandwidth for request $q$;

$n_q$, the number of assigned FSs to request $q$;

---

Algorithm I: Routing and spectrum assignment algorithm guaranteeing video quality

Given traffic $q$:
1. Run the *Dijkstra* algorithm and select $k$ shortest paths
2. *for every* candidate path $i \in [1:k]$:
   Determine the unoccupied FSs considering continuity and contiguity constraints, $D_i$
   *for every* candidate FSs set $j \in D_i$:
       Count the number of cuts on the connection links;
       Considering link establishment, calculate the misalignment change with neighbour links from source to destination nodes;
       *if* data type = *non-video*:
           $\beta = 0$
       Calculate the cost, $F_{i,j}$
   Select the local best FS set, $S_i = \max_{j \in D_i} F_{i,j}$
3. *if* $S_i = \emptyset$:
       go to Blocking
4. Select the best candidate path and FS set, $S = \max_{i \in [1:k]} S_i$
5. *if* $S$ did not satisfy expected video quality according to the application:
       go to Blocking
6. Establish connection
**Blocking:** block the request.

**Objective Function:**

*minimize*:

$$F_{\text{cost}}^q = \alpha \left[ N_c + \frac{N_m}{S_q N_{NL}} \right] + \beta \log \left( \frac{1}{U(BER_q)} \right)$$

The objective function minimizes the number of cuts and misalignments in EON network when applying RSA for a given connection request. This makes opportunity for the optical network to have more contiguous FSs to accommodate the coming connection requests in the future. Besides, it minimizes the cost of video, guaranteeing a minimum video quality and bring an optimal experience for the end user. The constraints of the proposed optimization problem are as follow.

**Constraints:**

1. $f_q - \frac{n_q}{2} \geq 0 \qquad \forall q \in Q$

2. $f_q + \frac{n_q}{2} \leq N_{FS} \qquad \forall q \in Q$

3. $f_{q'} - f_q > \frac{n_q}{2} + \frac{n_{q'}}{2}$
   $\forall l \in E : S_{l,q} \cdot S_{l,q'} \neq 0 \;,\; q' = \{Q - \{q\} | f_{q'} - f_q > 0\}$

4. $f_q - f_{q'} > \frac{n_q}{2} + \frac{n_{q'}}{2}$
   $\forall l \in E : S_{l,q} \cdot S_{l,q'} \neq 0 \;,\; q' = \{Q - \{q\} | f_{q'} - f_q < 0\}$

5. $S_{l,q}^{FS} = S_{l',q}^{FS}$
   $\forall l, l' \in E \;\&\; FS \in \left\{ f_q - \frac{n_q}{2} : f_q + \frac{n_q}{2} \right\}$

6. $U(BER_q) \geq U_{th} \qquad \forall q \in \{Q | \text{video data}\}$

In the above, the first and second constraints determine the lower and upper bounds for the usable bandwidth of fiber links, respectively. Moreover, jointly considering two first limitations, contiguity constraint will be satisfied. The constraints 3 and 4 ensure non-overlapping of bandwidth between any two connections $q$ and $q'$ with adjacent FSs. The continuity condition of FSs along connection links from source to destination is provided with constraint number 5. Finally, constraint number 6 guarantees a minimum quality of video for the end user. It should be noted that the later constraint applies only for video connection requests not for other data types.

## V. PERFORMANCE EVALUATION

### A. Simulation framework

To evaluate the proposed algorithm, we developed an event-based simulator. Considering a dynamic scenario, the lightpath requests appear randomly between source-destination pairs following *Poisson* process with arrival rate of $\lambda$, and each connection holding time follows a negative exponential distribution with parameter $\mu$. Moreover, the connection bandwidth is distributed between 1 and 10 FSs. The algorithm is tested on two benchmark network topologies, namely 14-node NSFNET and 24-node US Backbone, shown in Fig. 4. In our Monte Carlo simulations, we considered both video and non-video data connection requests so that, lightpath arrivals considered to be video streaming data by probability $p_v$ and non-video streaming data by probability $1 - p_v$. According to the presented statistics included in Introduction, it is wise to set $p_v = 0.8$.

### B. Metrics and Results

The offered load parameter is defined as $\frac{\lambda}{\mu}$ in Erlang. In the simulations, the traffic load varies within 100 to 600, where controlled by changing $\lambda$. One of the benchmark parameters for measuring the performance of RSA algorithms is blocking probability (BP). The BP is the ratio of blocked requests to the total connection requests, defined as follows

$$BP = \frac{\sum_{i=1}^{n_R} (N_i | \text{request } i = \text{blocked})}{\sum_{i=1}^{n_R} N_i} \qquad (11)$$

where $n_R$ is the total number of connection requests in the network for a given time and $N_i$ is the number of requested FSs for the $i$-th connection. In a dynamic scenario, the goal is to minimize the blocking probability. The second performance evaluation parameter is source-destination OSNR. Essentially, this parameter measures the quality of connection and obtains by averaging the $OSNR_{S-D}$, presented in (4), of all established lightpaths.

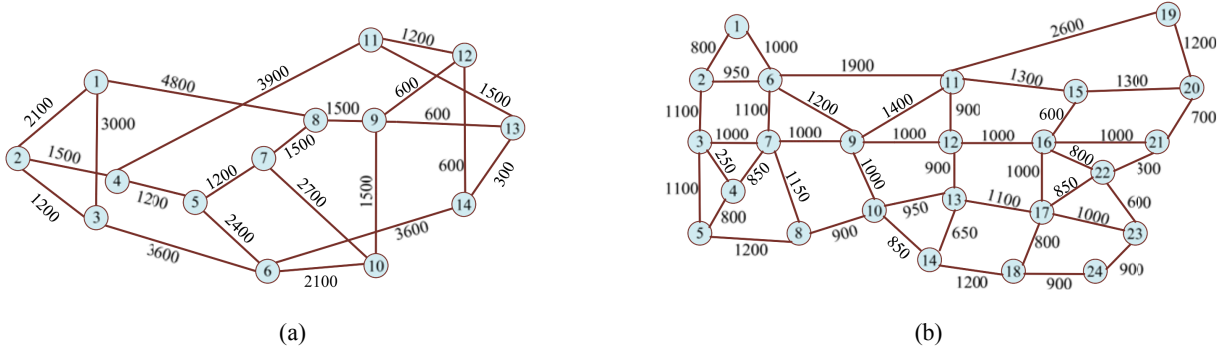

(a)　　　　　　　　　　　　　　　(b)

Fig. 4. Network topologies used to obtain numerical results. Fiber link lengths are in kilometers; a) NSFNET, b) US Backbone.

We have conducted simulations on two network topologies to evaluate the proposed dynamic RSA algorithm. The resultant mean OSNR is depicted in Fig. 5 for both 14-node NSFNET and 24-node US backbone topologies versus traffic load in Erlang. This result is evaluated separately for video and non-video data streams. As seen, in NSFNET topology the mean OSNR for video data is higher about 2dB in lower traffic loads and 1.3dB in higher traffic loads than mean OSNR for non-video. Likewise, for US backbone topology, the resultant mean OSNR for video data overcomes about 1.2dB the mean OSNR for non-video data. Comparing curves in Fig. 5 shows that for a given traffic load, the mean OSNR for US backbone is a little higher than OSNR for NSFNET. This is because that the mean link lengths in US backbone is lower than NSFNET and consequently, the fiber non-linearity effects are less.

The blocking probabilities versus traffic load in Erlang are shown in Fig. 6 for NSFNET and US backbone topologies. The BPs are separated for video and non-video data streams. As seen, the BPs for both data types are almost the same. This shows that in spite of an enhancement in OSNR, this algorithm not degrades the BP of video data.

The number of nodes and connection links in US backbone are more than in NSFNET. Consequently, as we use k-shortest path routing algorithm, under the same traffic load, the probability of finding a free path for a given connection request with adequate FSs in US backbone is higher than in NSFNET. Due to this, the BP in US backbone is lower than BP in NSFNET for equivalent traffic loads.

Fig. 7 compares the video decodable frame rate of proposed RSA and general RSA algorithms for NSFNET and US backbone topologies. The general RSA algorithm means that we do not distinguish between streaming data types, equivalently $\beta = 0$ in (7) for both data types. The illustrated results are obtained using resultant OSNR under some traffic loads. From the figure, one can see that for low traffic loads (high OSNRs) the difference between the DFR of proposed and general RSA algorithms is negligible, while the difference in higher traffic loads (low OSNRs) is high. This outcome and the values in charts give us a view that the functionality of DFR versus OSNR is concave down. Similarly, Fig. 8 compares the PSNR of decoded HEVC video using proposed and general

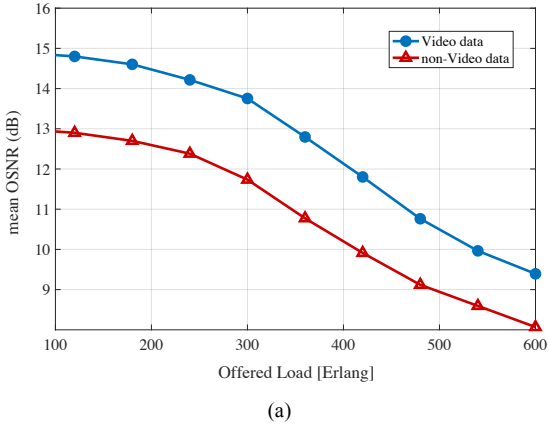 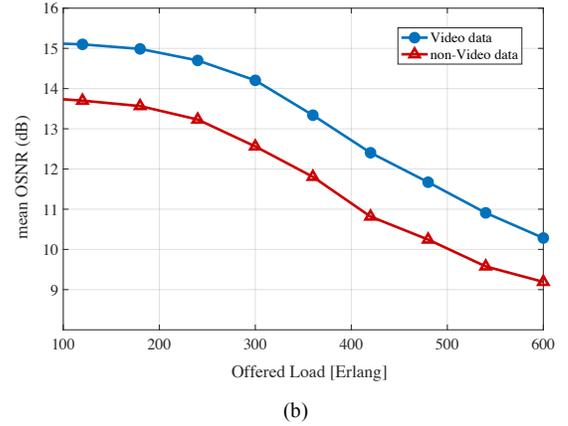

(a)            (b)

Fig. 5. OSNR vs. traffic load for video and non-video data connection requests; a) NSFNET b) US backbone.

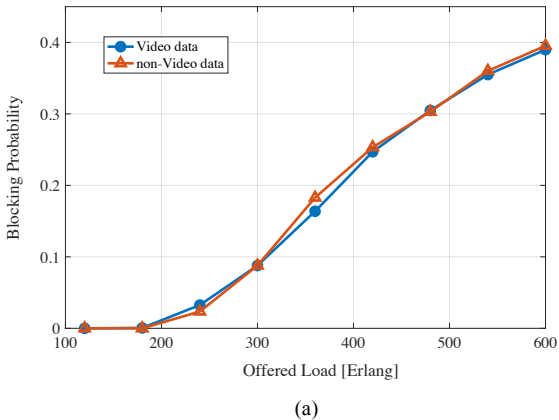 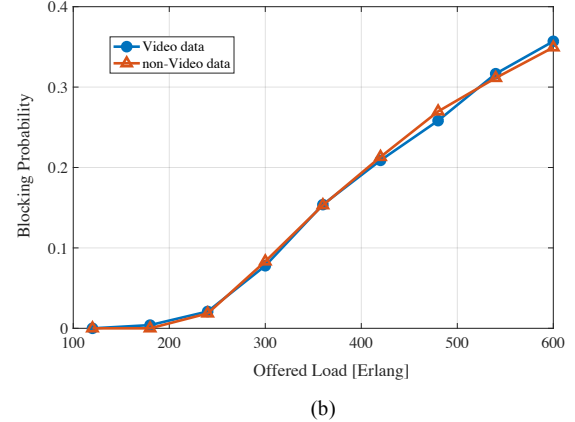

(a)            (b)

Fig. 6. Blocking probability vs. traffic load for video and non-video data connection requests; a) NSFNET b) US backbone.

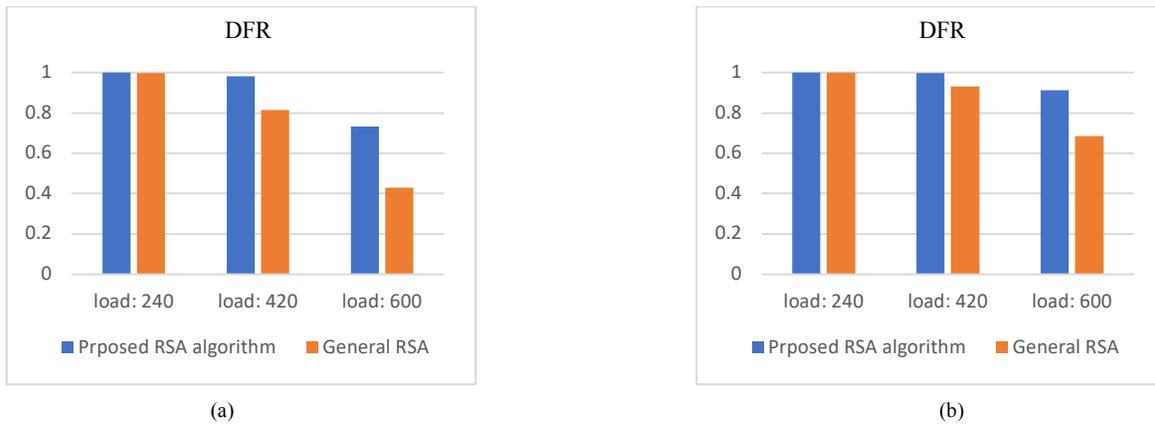

Fig. 7. Video decodable frame rate under some traffic loads for proposed and general RSA algorithms; a) NSFNET b) US backbone.

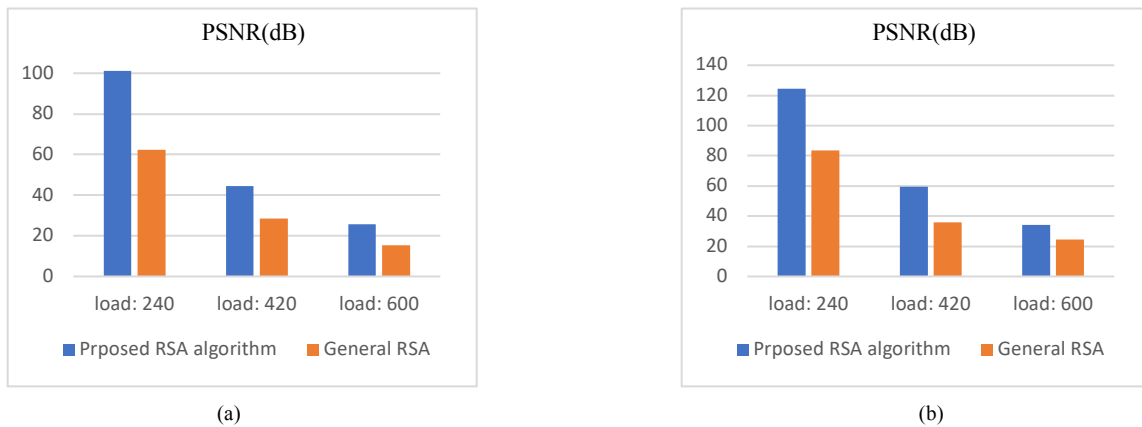

Fig. 8. Video PSNR under some traffic loads for proposed and general RSA algorithms; a) NSFNET b) US backbone.

RSA algorithms for NSFNET and US backbone topologies. Here, the difference between the PSNR of proposed and general RSA algorithms gets lower as traffic load increases (OSNR decreases).

According to the results, we can make an impression that, in spite of DFR, the functionality of DFR versus OSNR is concave up. Having this knowledge, one can tune the optimization parameters according to the application requirements in video streaming over EON.

## VI. Conclusion

In this paper, we focused on video communication over the elastic optical network. An innovative RSA algorithm is proposed in the form of an optimization problem, that the cost function is the weighted sum of network cost with a proposed video quality cost. Accordingly, in this problem, the constraints of network and video quality are considered jointly. The resultant sub-optimal answer gives routing path and assigned FSs indices for the given end to end demand where guarantees a minimum video quality and minimizes spectrum fragmentation. Using this RSA algorithm, the OSNR as an effective parameter in final video quality enters into the optimization problem. Accordingly, simulation results indicate that the mean OSNR of established connections with video content is higher around 1.5dB and 1.2dB than the OSNR of non-video connections for NSFNET and US backbone topologies, respectively.

**Hamed Alizadeh Ghazijahani** received his B.Sc., M.Sc., and Ph.D. degrees in telecommunication systems engineering from university of Tabriz, Iran in 2010, 2013, and 2019, respectively. He was visiting fellow with the Image, video and multimedia systems group at Singapore university of technology and design, in 2017. His main research interests include optical and wireless communication systems and multimedia communication.

**Hadi Seyedarabi** received his B.Sc. degree from University of Tabriz, Iran, in 1993, the M.Sc. degree from K.N.T. University of technology, Tehran, Iran in 1996 and his Ph.D. degree from University of Tabriz, Iran, in 2006 all in Electrical Engineering. He is currently an associate professor of Faculty of Electrical and Computer Engineering in University of Tabriz, Tabriz, Iran. His research interests are Image Processing, Computer Vision, Video Coding, Human-Computer Interaction, Facial Expression Recognition and Facial Animation.

**Javad Musevi Niya** was born in Ahar, Iran. He received his B.Sc. degree from the University of Tehran and his M.Sc. and Ph.D. Degrees both in communications from Sharif University of Technology (SUT) and the University of Tabriz, respectively. Since September 2006, he has been with the Faculty of Electrical and Computer Engineering of the University of Tabriz, where he is currently an Associate Professor. His current research interests include wireless communication systems, multimedia networks and signal processing for communication systems and networks.

**Ngai-Man Cheung** received the Ph.D. degree in electrical engineering from the University of Southern California, Los Angeles, CA, in 2008. He is currently an Associate Professor with the Singapore University of Technology and Design (SUTD). From 2009-2011, he was a postdoctoral researcher with the Image, Video and Multimedia Systems group at Stanford University, Stanford, CA. He has also held research positions with Texas Instruments Research Center Japan, Nokia Research Center, IBM T. J. Watson Research Center, HP Labs Japan, Hong Kong University of Science and Technology (HKUST), and Mitsubishi Electric Research Labs (MERL). His work has resulted in 10 U.S. patents granted with several pending. His research interests include signal, image, and video processing.